# MULTI-BAND DIPOLE AND MULTIPOLE WAKEFIELDS IN NLC TRAVELING WAVE ACCELERATORS USING A WIRE MEASUREMENT TECHNIQUE

R.M. Jones[†], N. Baboi[†]*, and N.M. Kroll[§]

[†]Stanford Linear Accelerator Center,
2575 Sand Hill Road, Menlo Park, CA, 94025
[§]University of California, San Diego
La Jolla, CA 92093-0319
*On leave from NILPRP, P.O. Box MG-36, 76900 Bucharest, Romania

## Abstract

Dipole wakefields in NLC (Next Linear Collider) structures have been measured with ASSET [1] and well predicted by a circuit model [2]. However, the experimental technique is both time-consuming and expensive. Here, we report on kick factor and synchronous frequency determination for $1^{st}$ and higher order dipole bands for TW (Traveling Wave) accelerators via a wire measurement technique. This stand-alone technique is relatively inexpensive and may lead to an efficient determination of wakefield parameters. The perturbative effect of the wire on the dipole band is pointed out and a two-wire scheme with a limited perturbative effect is also discussed.

*Paper presented at the 2002 8$^{th}$ European Particle Accelerator Conference (EPAC 2002)
Paris, France,
June 3$^{rd}$ -June 7$^{th}$, 2002*

This work is supported by Department of Energy grant number DE-AC03-76SF00515† and DE-FG03-93ER40759§

# MULTI-BAND DIPOLE AND MULTIPOLE WAKEFIELDS IN NLC TRAVELING WAVE ACCELERATORS USING A WIRE MEASUREMENT TECHNIQUE

R.M. Jones[†], N. Baboi[†]*; SLAC, Stanford, CA 94309, N.M. Kroll[§]; UCSD, La Jolla, CA

*Abstract*

Dipole wakefields in NLC (Next Linear Collider) structures have been measured with ASSET [1] and well predicted by a circuit model [2]. However, the experimental technique is both time-consuming and expensive. Here, we report on kick factor and synchronous frequency determination for 1st and higher order dipole bands for TW (Traveling Wave) accelerators via a wire measurement technique. This stand-alone technique is relatively inexpensive and may lead to an efficient determination of wakefield parameters. The perturbative effect of the wire on the dipole band is pointed out and a two-wire scheme with a limited perturbative effect is also discussed.

## 1. INTRODUCTION

In the NLC we plan to accelerate 192 bunches each of which consists of the order of $10^{10}$ particles. This bunch train is accelerated within a single RF pulse. As the bunches are highly energetic, it is clear that a non-negligible wakefield will be left behind each accelerated bunch. This wake can at the very least considerably dilute the final emittance of the beam or, at worst can give rise to a BBU (Beam Break Up) instability [3] causing trailing bunches to oscillate transversely with increasing amplitude. For this reason, it is important to be able to predict the transverse wakefields that will be excited in a given accelerating structure. The wakefield can be decomposed in various multipoles, and for the structures under consideration for the NLC the dipole wakefield is considered to be the most serious component. This dipole field can in itself be decomposed into a band structure. The most serious mode we have found to be located in the 1st band, nonetheless, as will be shown in the following sections, the 3rd and 6th bands have significant contributions to the overall wakefield for the TW structures already fabricated at SLAC and KEK [4].

A vigorous program of investigating SW (Standing Wave) accelerators for the NLC is also in progress at SLAC, motivated by significantly reduced breakdown events being observed at high gradients (~70MV/m) compared to their TW counterparts. However, the partitioning of the wakefield amongst the dipole bands is strikingly different from that in TW structures [5]. All modes up to the 3rd band have almost equal contributions to the wakefields and consequently attention will be paid to damping all of these modes.

We have been able to successfully predict these wakefields using a spectral function [2] technique which is able to incorporate the essential physics of the accelerating structures and to include any experimental errors into the simulation. We have compared these theoretical predictions with experiments to measure the wakefield using the ASSET facility at the SLC (Stanford Linear Collider). ASSET consists of a sector incorporated into part of the SLC to measure the transverse deflection on an electron bunch caused by a positron drive bunch and from this the wakefield is derived. However, this experiment is both time consuming and it requires dedicated beam time to be allocated from a national laboratory's program.

Thus, a stand-alone bench measurement of the wakefield is clearly highly desirable from the standpoint of being able to routinely measure the wakefield in a series of structures and in order to make this measurement in a relatively inexpensive manner. A wire measurement method to simulate the wakefield left behind a particle by the progress of an RF pulse on a wire is already in the process of being set up at SLAC [6]. In this paper, we discuss the band structure likely to be encountered in a typical cell from an accelerating structure and on coupling these modes to the wire. We also point out the possible advantages of a two-wire method. From this information we intend to develop a circuit model to describe the couplings to a multi-cell accelerating structure and this will be the subject of a future publication.

## 2. BAND STRUCTURE AND COUPLING IN WIRE MEASUREMENT

The first experiment at SLAC that we intend to do using the wire measurement technique will be on DS2 (Detuned Structure number 2), which has been fabricated several years ago and is readily available for testing. We will conduct this experiment on the last 50 cells of the 206 cell structure. In order to understand the mode coupling to the wire we chose to model cell 198, which is the last non-special cell [7] of the structure. We used the finite eigenmode solver in the commercially available code, HFSS (High Frequency Structure Simulator), to model the dispersion characteristics of three separate cases: a single cell loaded with no wire at all, a single wire loading the structure at various offsets from the electrical centre and two wires symmetrically displaced from the center of the cavity. To verify these results we have used the finite difference electromagnetics code GdfidL [8]

The effect of loading the structure with a single wire can be found by comparison of Figs. 1a and 1b which shows, as one would expect, that a fine wire in the cell center has little effect on the dipole mode and almost no effect on higher order multipoles. While it does add a TEM-coax-like mode to the spectrum, there is a very strong coupling between that mode and all the TM monopole modes. The light line indicated on all curves, corresponds to a mode propagating in free-space and also a mode with a phase velocity of a pure coaxial TEM mode. Every frequency at which the light line crosses a TM monopole shows a very strong avoided crossing effect with the putative coax mode. The effect is so strong that the light-line character of the coax is only occasionally discernable.

[†] Supported by DOE grant number DE-AC03-76SF00515
[§] Supported by DOE grant number DE-FG03-93ER40759
*On leave from NILPRP, P.O. Box MG-36, 76900 Bucharest, Romania

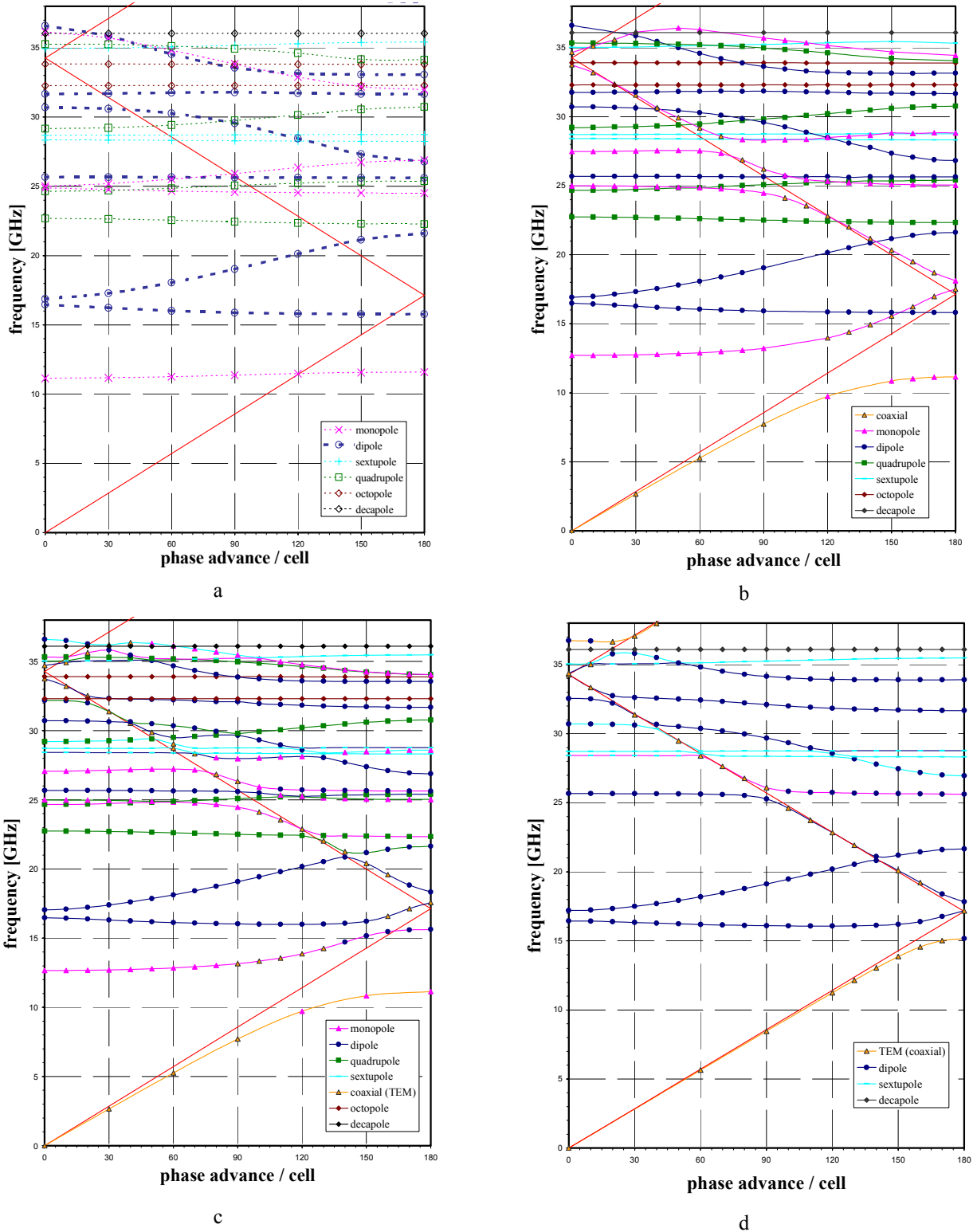

Figure 1 Dispersion diagrams for cell 198 of DS2 with periodic boundary conditions. The cavity walls are perfect conductors: a) no wire is present; a quarter of cell is simulated with both symmetry planes having magnetic boundary conditions combined with one having magnetic and the other electric boundary; b) a wire with diameter of 250 µm is stretched in the center of the cavity; half of a cell is simulated, with magnetic boundary conditions in the symmetry plane; c) a half cell as in b) with the wire displaced by 1 mm along the symmetry plane; d) two wires 250 µm in diameter are placed symmetrically around the cavity axis 2 mm apart; a quarter of a cell is simulated, with the wire placed in one symmetry plane having magnetic boundary conditions, while the other plane has an electric boundary.

It is helpful to next consider Fig. 1d, which consists of two wires, each of which are displaced 1mm from the axis of the cavity. While the wire pair has two TEM-like modes, presumably one would drive only the twinax mode, a mode which couples only to the odd multipoles and whose light-line character is evidently well preserved. The avoided crossing effect is discernable wherever the light-line crosses an odd multipole. The coupling is strongest for the first, third, and sixth dipoles, but weaker for the second and higher order multipole bands. We note however, that in the initial experiment that will be performed at SLAC only a single wire will be used although driving the twinax modes may be an option at some later stage. Thus qualitatively, at least, it follows the behavior of the kick factors, and apart from practical problems twinax excitation shows high promise as a transverse wake simulator. Fig. 1c shows the effect of displacing a single wire off center, which would presumably be more practical than the twinax method. While this procedure mixes all the multipoles to some extent, the strongest avoided crossing-like effects involve the crossing of the dipoles with the coax-monopole hybrids, crossings which, can be quite remote from light-line crossings. These plots provide an initial orientation for interpretation and analysis of forthcoming experimental studies.

## 3. TRANSVERSE KICK FACTOR AND WAKEFIELD ANALYSIS

The transverse wakefield excited by a particle beam can be decomposed into modes which kick the beam transversely to the axis of acceleration. Here, we use a quasi-coupled analysis in which we calculate the wakefield at the synchronous frequency using the individual cell kick factor. The envelope of the wakefield is written as the absolute value of a summation:

$$W(s) = 2\left|\sum_{n=1}^{N} K_n \exp\left[j\frac{\omega_n s}{c}\left(1+\frac{j}{2Q_n}\right)\right]\right| \quad (2.1)$$

where s is the distance behind the bunch, $K_n$ is the transverse kick factor, $\omega_n/2\pi$ is the synchronous frequency, $Q_n$ is the quality factor of the mode (which is of course infinite when no damping is present) and the subscript n refers to the particular cell. The kick factor is evaluated as:

$$K_n = \frac{c}{4\omega_n U_n \Delta r^2 L}\left|\int_L E_z(z,\Delta r)\exp[j\omega_n s/c]dz\right|^2 \quad (2.2)$$

where $E_z(z,\Delta r)$ is the electric field at a displacement $\Delta r$ from the axis, $U_n$ is the energy stored in the mode. The integral is performed over the length L of the cavity. The kick factor for lowest dipole mode is shown in Fig 2., evaluated at two transverse offsets for all three cases. In the limit of zero offset of the wire both the single and dual wire cases are seen to approach that of the structure without the wire.

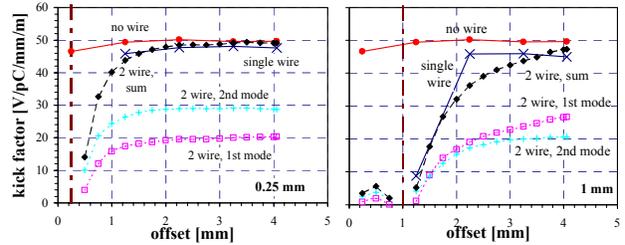

Figure 2: Kick factors of cell 198 of the first dipole synchronous mode as function of the offset path of integration for two wire offsets: 0.25mm in the leftmost and 1mm in the rightmost.

We also evaluated the kick factor for 5 further bands and several further cells. These are displayed in Fig 3 together with the wakefield that results from each, undamped, individual band for the complete 206 cell structure. The first band clearly is the most serious. However, the beam dynamics simulations have indicated that BBU is likely to occur for wakefields greater than 1V/pC/mm/m and for both the $3^{rd}$ and $6^{th}$ bands this is seen to be the case.

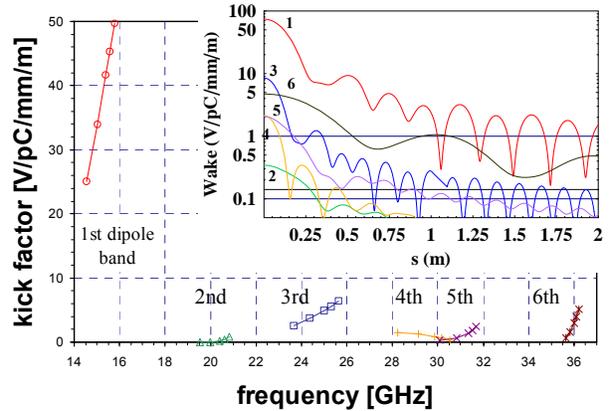

Figure 3: Transverse kick factors for six dipole bands of DS2. Shown inset is the envelope of the wakefield.